\documentclass[aps,pre,reprint,superscriptaddress]{revtex4-1}
\usepackage[dvipdfmx]{graphicx}
\usepackage{amssymb,amsmath}
\usepackage{bm}
\usepackage{color}
\newcommand{\rh}[1]{\rho_{\rm{#1}}}

\begin{document}


\title{Taming Macroscopic Jamming in Transportation Networks}


\author{Takahiro Ezaki}
\email{ezaki@jamology.rcast.u-tokyo.ac.jp}
\affiliation{Department of Aeronautics and Astronautics, Graduate School of Engineering, The University of Tokyo, 7-3-1  Hongo, Bunkyo-ku, Tokyo 113-8656, Japan}
\affiliation{Japan Society for the Promotion of Science, 8 Ichibancho, Kojimachi, Chiyoda-ku, Tokyo 102-8472, Japan}

\author{Ryosuke Nishi}
\altaffiliation[Present address: ]{Department of Mechanical and Aerospace Engineering, Graduate School of Engineering, Tottori University, 4-101 Minami, Koyama, Tottori 680-8552, Japan}
\affiliation{National Institute of Informatics, 2-1-2 Hitotsubashi, Chiyoda-ku, Tokyo 101-8430, Japan}
\affiliation{JST, ERATO, Kawarabayashi Large Graph Project, 2-1-2 Hitotsubashi, Chiyoda-ku, Tokyo 101-8430, Japan}

\author{Katsuhiro Nishinari}
\affiliation{Department of Aeronautics and Astronautics, Graduate School of Engineering, The University of Tokyo, 7-3-1 Hongo, Bunkyo-ku, Tokyo 113-8656, Japan}
\affiliation{Research Center for Advanced Science and Technology, The University of Tokyo, 4-6-1 Komaba, Meguro-ku, Tokyo 153-8904, Japan}


\date{\today}

\begin{abstract}
In transportation networks, a spontaneous jamming transition is often observed, e.g. in urban road networks and airport networks. Because of this instability, 
flow distribution is significantly imbalanced on a macroscopic level. 
To mitigate the congestion, we consider a simple control method, in which 
congested nodes are closed temporarily, and investigate how it influences the overall system. Depending on the timing of the node closure and opening, and congestion level of a network, the system displays three different phases: free-flow phase, controlled phase, and deadlock phase.
We show that when the system is in the controlled phase, the average flow is significantly improved, whereas
when in the deadlock phase, the flow drops to zero. We study how the control method increases the network flow and obtain their transition boundary analytically. 
\end{abstract}


\maketitle

\section{introduction}
Transportation on networks plays an essentially prerequisite role in our social activities. 
Complex network structures enable goods supply \cite{Radon, Helbing, HelbingL}, information communication \cite{Alnet, Huberman}, and human mobility \cite{Daganzo1,Daganzo2,Daganzo3,Esser,HelbingLammerLebacque,Gugat,Li, Guimera1, Guimera2, Zanin, Bagler} in a vast human society. 
However, as a drawback of this efficiency, the structural complexity often causes breakdowns of the systems \cite{DaganzoUrbanGridlock,Mahmassani,Zheng,Simonsen}. 
In particular, traffic networks (e.g. air traffic networks \cite{EzAir} and  road traffic networks \cite{Daganzo3}) 
display a macroscopic jamming transition when the system congestion exceeds a certain level.
It is essentially different from the classically known (microscopic) jamming phenomenon \cite{HelRev,NagaRev} and appears 
only when subsystems (e.g., roads or airports) are connected with each other. 
This macroscopic jamming is believed to explain (at least partially) severe jamming phenomena in reality \cite{EzAir,Daganzo3}.
A simple guaranteed solution to the problem is to increase the effective capacity of the system,
which is, however, not always feasible in our society confronted by rapidly increasing traffic demand. 
Hence, some external control method is necessary to mitigate the congestion. 

In such complex systems, understanding how automatic controls affect their macroscopic features 
still remains a grand challenge; nevertheless we have to deal with such exigent problems in reality.
\textcolor{black}{To facilitate the understanding of network traffic under control, innumerable studies have
been conducted (see e.g., Ref. \cite{Papageorgiou} and references therein). 
However, because until very recently the issue of the macroscopic jamming transition on networks has not been explicitly discussed, few studies have focused on the outcome of control to mitigate such jamming.} 
In this paper, we consider simple controls, which close nodes adaptively by temporarily removing links leading to congested nodes and reconnecting them when the congestion is resolved.
This type of simple on-off controls is widely adopted, e.g. in traffic light controls \cite{Papageorgiou, LammerHelbing2008,SchadschneiderChowdhuryNishinari} and ramp metering \cite{PapageorgiouKotsialos,Papageorgiou} for vehicular traffic, and the ground delay programs \cite{GDP} for air traffic \cite{Luo,Rosenberger,Lan}.
We show that this control method could cause a sudden breakdown, which is characterized by a discontinuous
phase transition to the \textit{deadlock} state, whereas it is partly effective for alleviating congestion in a certain parameter regime. This deadlock transition stems from the control we adopt, and is a different phenomenon from previously known deadlock phenomena observed e.g. in the BML model \cite{BML, Tadaki, Tadaki2}. 
By disclosing the dynamics caused by the control, we discuss its effectiveness and 
criteria for designing a control law on transportation systems.

\begin{figure*}[tbhp]
\includegraphics[width=100mm]{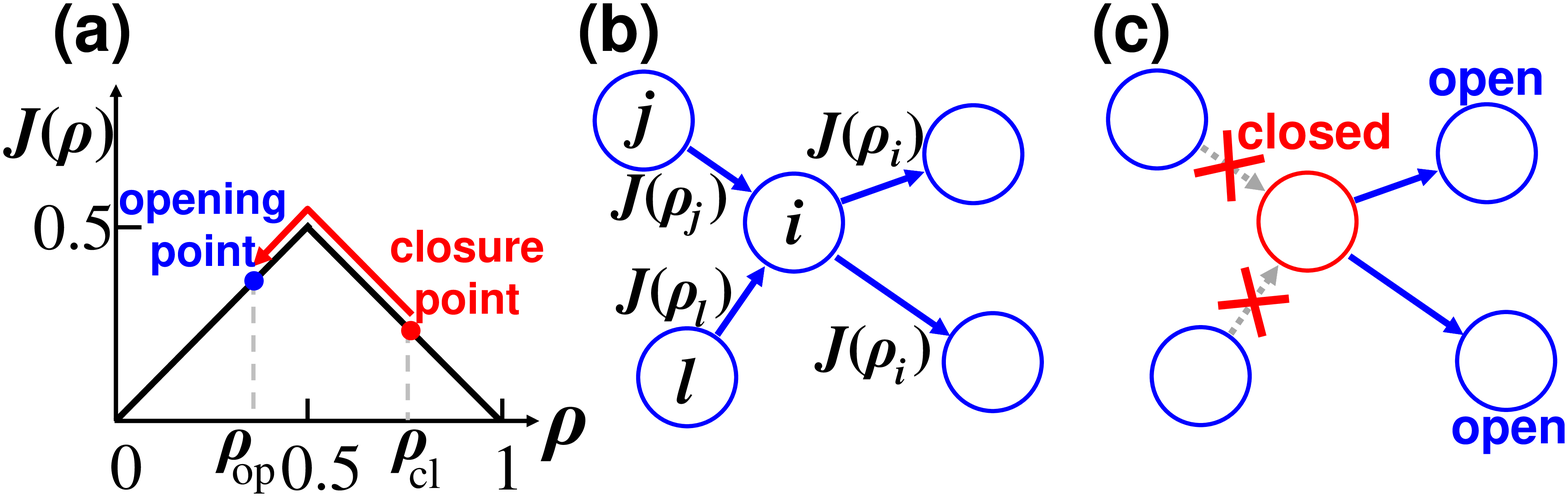}
\caption{ (a) Flow-density relationship and control points. (b) Inflow to and outflow from a node in the network. (c) Disconnection of a link leading to a congested node.}
\label{rule}
\end{figure*} 

\begin{figure}[tbhp]
\includegraphics[width=90mm]{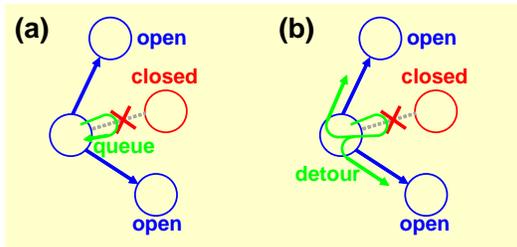}
\caption{\textcolor{black}{(a) Queuing rule. (b) Detouring rule.}}
\label{qdrule}
\end{figure}

\begin{figure*}[t]
\includegraphics[width=170mm]{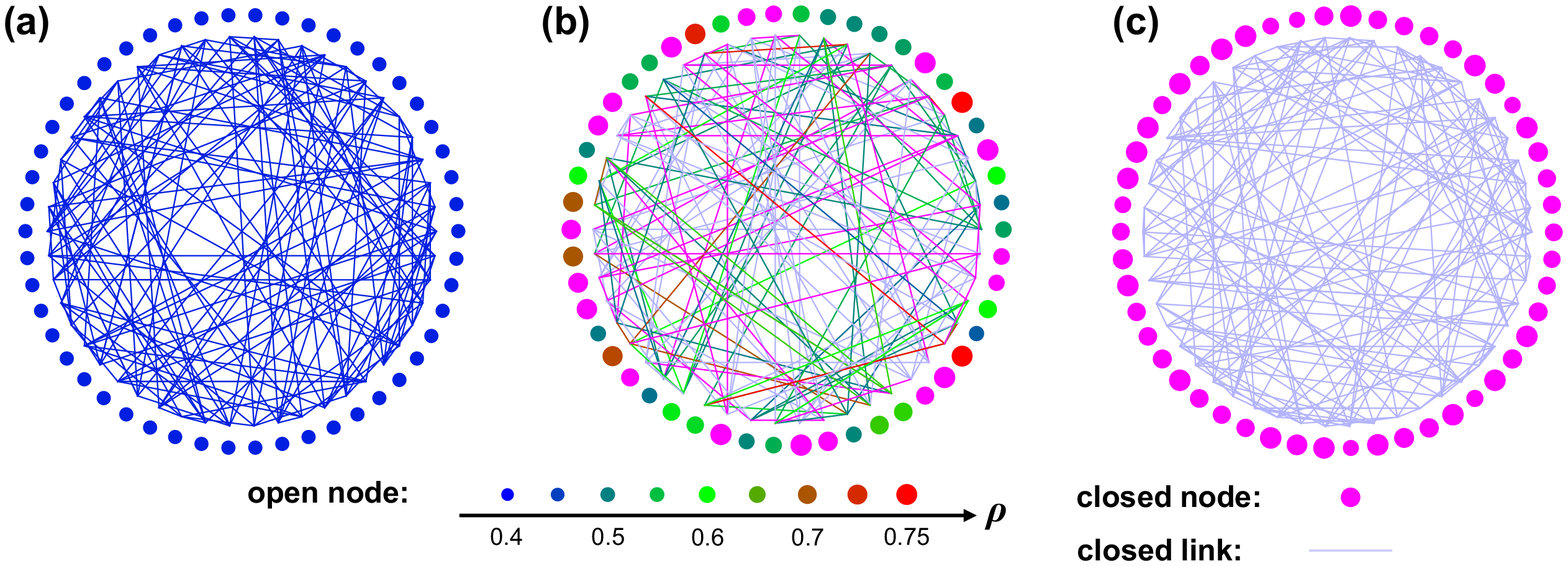}
\caption{ Snapshots of (a) the free-flow phase ($\bar{\rho}=0.4$), (b) the controlled phase  ($\bar{\rho}=0.55$) and (c) the deadlock phase  ($\bar{\rho}=0.65$). Each disk represents a node, whose radius and color correspond to the density value. When a node is closed, the node and disconnected links are expressed by a magenta disk and gray lines, respectively. For display purpose, the networks are generated for $N=50$ and $k=3$.}
\label{sst}
\end{figure*} 
\section{Model}
We begin by reviewing the original model \cite{Daganzo1, Daganzo2, Daganzo3, EzAir} of network transportation without controlling rules.
It is known that flow of spatially exclusive objects often nonlinearly depends on the density 
of the objects \cite{SchadschneiderChowdhuryNishinari,HelRev, NagaRev}. The flow-density relationship consists of two parts; 
i.e., free-flow and jamming regions, which are characterized by the increase and decrease of
flow according to density, respectively.
We consider density distribution on a network comprised of $N$ nodes that are labeled by $i$ (, $j$ and $l$)$=1,\cdots N$. 
For each density value in node $i$, $\rho_i$, flow on a link from the node is determined by the function value, $J(\rho_i)$.
For simplicity, we assume a piecewise linear flow-density relationship [Fig. \ref{rule}(a)], $J(\rho_i)=\min{\{\rho_i,1-\rho_i\}}$.
Each node sends and receives density through the links [Fig. \ref{rule}(b)] defined by an adjacency matrix $A$, 
whose element $A_{ij}$ is $1$ if there is a link from node $i$ to $j$, and $0$ otherwise.
Hence, time development of the density in node $i$ is written as 

\begin{equation}
\frac{d \rho_i}{dt} = \sum_{j=1}^{N} A_{ji}J(\rho_j) - \sum_{j=1}^{N} A_{ij}J(\rho_i).\label{me}
\end{equation}

In this paper, since we are interested in the nature of the system under control, 
we ignore individual properties of nodes; the flow-density relationship, capacity and degree of nodes.
Moreover, for simplicity directed random networks with homogeneous in- and out-degrees $k_{\rm{in}}=k_{\rm{out}}=k$ are used.

It is known that this model [Eq. (\ref{me})] exhibits spontaneous destabilization for large densities \cite{Daganzo3,EzAir}.
When the density in a node is in the jamming regime ($\rho>\frac{1}{2}$), a small increase in density leads to a decrease of outflow, which causes further congestion. This effect destabilizes uniform flow in networks (for a more mathematical explanation, see Appendix \ref{stabcon}). 
Hence, we consider a control method as follows.
Since when the density in the system is high, density concentrates in congested nodes,
the method must disperse it in some way.

As the simplest control, we here consider the following operations:
(i) when the density in node $i$ exceeds $\rho_{\rm{cl}}$, links leading to the node are disconnected [Fig. \ref{rule}(c)], i.e., 
$A_{ji}=1\rightarrow A_{ji}=0$ ($j\neq i$) ($closed$);
(ii) when the density in a closed node drops below $\rho_{\rm{op}}$, the incoming links are all reconnected [see also Fig. \ref{rule}(a)].
The flow that is supposed to travel through the disconnected links must stay in the present node 
or detour using other links. In this paper, both rules are studied separately. 
The first rule is the \textit{queuing rule} \textcolor{black}{[Fig. \ref{qdrule}(a)]}, wherein the canceled flow stays in the departure node, 
resulting in no change in the system equation for open nodes (\ref{me}), whereas for closed nodes the inflow
term vanishes:
\begin{equation}
\frac{d \rho_i}{dt} =\left\{ \begin{array}{ll}
 \sum_{j=1}^{N} A_{ji}J(\rho_j)  -  \sum_{j=1}^{N} A_{ij}J(\rho_i)  & \qquad (i:\rm{open}),\\
- \sum_{j=1}^{N} A_{ij}J(\rho_i)  .\label{cl} & \qquad (i:\rm{closed}).\\
\end{array} \right.
\end{equation}
 
The second rule is the \textit{detouring rule} \textcolor{black}{[Fig. \ref{qdrule}(b)]}, which redistributes the canceled flow equally to open links.
Thus, flux to an open node is increased by this  additional flow:
\begin{equation}
\frac{d \rho_i}{dt} =
\left\{ \begin{array}{ll}
\sum_{j=1}^N\frac{k J(\rho_j) }{\sum_{l=1}^{N} A_{jl}\sigma (l)}- kJ(\rho_i) & \qquad (i:\rm{open}),\label{detour} \\
-kJ(\rho_i) & \qquad (i:\rm{closed}),\\
\end{array} \right.
\end{equation}
where $\sigma(l)$ represents the state of node $l$, i.e., open ($\sigma(l)=1$) and closed ($\sigma(l)=0$) states. For closed nodes, the equation is 
not influenced by the state of neighboring nodes if at least one outgoing link is open. 
When the destination nodes are all closed, the outflow from a node vanishes, regardless of its state.

\subsection{Equivalence of the two control rules}
In a mean-field treatment (large $N$ and $k$; see Appendix \ref{equi}), the system equations of 
the two rules are equivalent through a transformation, $\frac{t_{\rm{q}}}{x}=t_{\rm{d}}$, where $x=\frac{\sum_{i=1}^{N}{\sigma(i)}}{N}$ is the ratio of open nodes in the 
thermodynamic limit. 
Hence, both rules result in the identical density distribution in the mean-field limit.

\subsection{Parameter settings}
In this paper we restrict ourselves to considering the case of $\rho_{\rm{cl}}=0.75$,
and focus on the system's dependence on $\rho_{\rm{op}}$ and the average density, $\bar{\rho}=\frac{\sum_{i=1}^N{\rho_i}}{N}$.
We conducted simulations for $N=100$ and $k=10$ to compare the results with the predictions obtained by a mean-field theory. 
Note that the arguments in this paper are not affected significantly by these specifications. (See also Appendix \ref{infk}.)
The system equations were integrated by using the Runge-Kutta fourth-order method with $dt = 0.00 01$. 

\section{Results}
\subsection{Free-flow, controlled and deadlock phases}
First, we focus on the case of $\rho_{\rm{op}}=0.5.$
Depending on the value of the average density, 
the system exhibits three different phases, i.e., the free-flow, controlled, and deadlock phases (Fig. \ref{sst}).
These phases are observed in the system both under queuing and detouring operations.
When the average density is small, congested nodes do not appear 
and uniformly distributed flow is achieved [Fig. \ref{sst}(a)].
As the average density increases, congested nodes appear  
and our control rules apply. Because these operations increase the load in neighboring nodes, congested nodes appear on different locations while the operated node is recovering from the congested state. 
Thus, a certain ratio of nodes are almost constantly closed; such a state is referred to as the controlled phase.  
When we further increase the average density, 
all the nodes become closed. In this state, nodes cannot accept nor send density, and thus flow in the system vanishes.
Since the node capacity is temporarily limited during node closure, the operations effectively reduce the total capacity of the system.

\begin{figure}[tbhp]
\includegraphics[width=80mm]{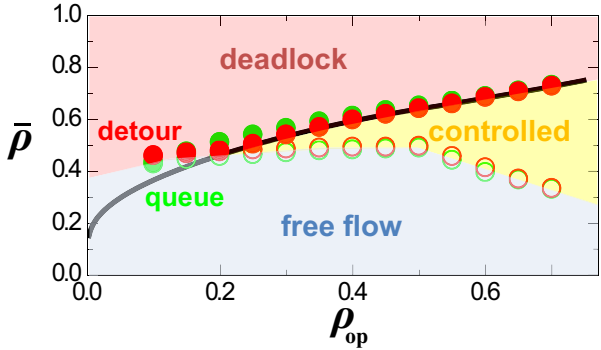}
\caption{ Phase diagram of the system. The closed and open circles represent transition points 
obtained by simulations. The black bold line is the theoretical prediction, Eq. (\ref{tl}). 
The simulation results were obtained by examining the presence of a closed node after a sufficiently long time ($t=100$). At $t=0$, $10$ nodes were selected randomly to be closed with $\rho=0.75$. 
With this initial condition, the system does not accidentally fall into the free flow (absorbing) state for system parameters in the other two phases.}
\label{pd}
\end{figure}

Next, we discuss the phase transitions for various $\rho_{\rm{op}}$ values.
Fig.  \ref{pd} shows the phase diagram in the $\bar{\rho}$--$\rho_{\rm{op}}$ plane.
The simulation results were obtained by observing the state of the system after $10^6$ simulation time steps. 
The initial density distribution in the system was set uniformly, but perturbed with randomly selected $10$ closed nodes ($\rho=\rho_{\rm{cl}}$). 
When $\rho_{\rm{op}}$ is small, the controlled phase is not observed.
In this region, closure time is too large, and thus it adversely affects the system by utilizing less of the capacity of the network. For large $\rho_{\rm{op}}$ values, the controlled phase exists  
when the average density is moderately large. Here the control operations function without causing a breakdown. 
As a result, interestingly, the phase diagram has a triple point, which reflects the limit point of controllability. 
Because large $\rho_{\rm{op}}$ values lead to small closure time of controlled nodes, the controlled phase region expands for such  $\rho_{\rm{op}}$ values. 

In a practical context, the transition line from the controlled phase to the deadlock phase 
is important to prevent a breakdown.  Since the two rules are equivalent in the mean-field limit, 
we analyze the detouring rule, in which the time development of density in a closed node is known [Eq.(\ref{detour})].
First, closed nodes in the controlled phase are focused on.
We assume that the probability distribution of density in the time development, $Q(\rho)d\rho$, is proportional to 
the time in which the density is in the interval $[\rho, \rho+d\rho]$ in the closed state, i.e., $Q(\rho)\propto\left|\frac{dt}{d\rho}\right|.$ 
From Eq. (\ref{detour}), we obtain $\left|\frac{dt}{d\rho}\right|=\frac{1}{kJ(\rho)}.$
Thus, the distribution is finally written as
\begin{equation}
Q(\rho) = T_{\rm{cl}\rightarrow \rm{op}}^{-1} \frac{1}{kJ(\rho)},
\end{equation}
where the normalization constant, $T_{\rm{cl}\rightarrow \rm{op}}$, corresponds to the time 
from closure to the next opening of a controlled node:
\begin{eqnarray}
T_{\rm{cl}\rightarrow \rm{op}} &=& \int_{\rho_{\rm{op}}}^{\rho_{\rm{cl}}} \frac{d\rho}{kJ(\rho)}\nonumber\\
&=&\left\{ \begin{array}{ll}
\int_{\rho_{\rm{op}}}^{\frac{1}{2}} \frac{d\rho}{k\rho} + \int_{\frac{1}{2}}^{\rho_{\rm{cl}}} \frac{d\rho}{k(1-\rho)}    &\qquad (\rho_{\rm{op}} < \frac{1}{2}), \nonumber\\
\int_{\rho_{\rm{op}}}^{\rho_{\rm{cl}}} \frac{1}{k(1-\rho)} &\qquad (\rho_{\rm{op}} \geq \frac{1}{2}), \\
\end{array} \right.\\
&=&\left\{ \begin{array}{ll}
\frac{1}{k}\log{\frac{1}{4(1-\rho_{\rm{cl}})\rho_{\rm{op}}}}    & \qquad (\rho_{\rm{op}} < \frac{1}{2}), \\
\frac{1}{k} \log{\frac{1-\rho_{\rm{op}}}{1-\rho_{\rm{cl}}}}    & \qquad (\rho_{\rm{op}} \geq \frac{1}{2}). \\
\end{array} \right.
\end{eqnarray}
Using this distribution, the average density in closed nodes is calculated as 
\begin{eqnarray}
\bar{\rho}_{\rm{closed}} &=& \int_{\rh{op}}^{\rh{cl}} \rho Q(\rho)d\rho \nonumber\\
&=&\left\{ \begin{array}{ll}
\frac{\rh{cl}+\rh{op} + \log{2(1-\rh{cl})} -1}{\log{4(1-\rho_{\rm{cl}})\rho_{\rm{op}}}} & (\rho_{\rm{op}}<\frac{1}{2}), \\
1 - \frac{\rh{cl}-\rh{op}}{\log{\frac{1-\rho_{\rm{op}}}{1-\rho_{\rm{cl}}}}}. & (\rho_{\rm{op}}\geq \frac{1}{2}). \label{tl}\\
\end{array} \right.
\end{eqnarray}
At the transition point when the deadlock occurs (all the nodes become closed), this value should coincide with the overall average density, i.e., $\bar{\rho}\simeq\bar{\rho}_{\rm{closed}}$.  
This expression is shown in Fig. \ref{pd}, and gives an accurate prediction when $\rho_{\rm{op}}$ is large.
Because this approximate analysis assumes that the number of closed nodes continuously increases
from the controlled phase to the deadlock phase with increasing average density, the equation cannot describe the transition 
from the free-flow phase to the deadlock phase for small $\rho_{\rm{op}}$. 
The agreement between this equation and the simulation results implies that 
the closure operations can be regarded to occur almost at random without causing synchronization.

\subsection{Macroscopic flow-density relationship}
\begin{figure}[tbhp]
\includegraphics[width=80mm]{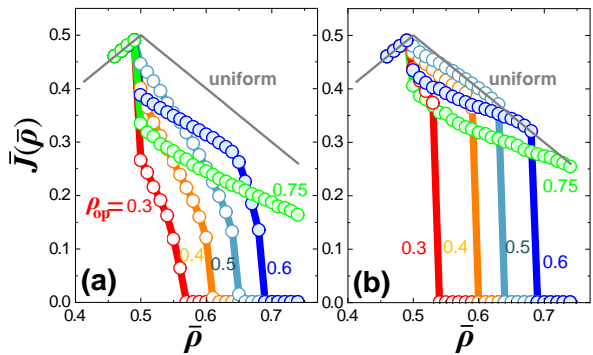}
\caption{Macroscopic flow-density relationship for (a) the queuing rule and (b) the detouring rule. Note that, here the initial density values are set as $\rho_i = \bar{\rho} + \delta_i$, where $\delta_i$ is a small random value drawn from a uniform distribution $[-0.005,0.005]$.}
\label{fds}
\end{figure}

As the most important performance index of the system, we study a macroscopic relationship between the average flow per link and the average density, which is defined by $\bar{J}=\frac{\sum_{i=1}^N\sum_{j=1}^{N}A_{ij}J(\rho_i)}{Nk}$. Figures \ref{fds}(a) and \ref{fds}(b) show the  macroscopic flow-density relationship for the queuing rule and detouring rule, respectively.
It is convenient for understanding the results to consider the base line case $\rho_{\rm{op}} = \rho_{\rm{cl}} =0.75$, wherein nodes reject inflow only when their (prescribed  permissible) capacity ($\rho_{\rm{cl}}$) is fully occupied. 
Because of the system's intrinsic instability, the uniform density distribution is not achieved in the density region $\bar{\rho}>\frac{1}{2}$, in which the macroscopic flow decreases.  
If the macroscopic flow is larger than that in the base line case, the controls are effective. In some cases, we can partly see the effectiveness in the controlled phase; however, when the average density further increases,
the macroscopic flow sharply drops to zero, which corresponds to the phase transition to the deadlock phase. 
When $\rho_{\rm{op}}$ is small, the closure time is large, and thus the system falls into the deadlock phase even with a low average density, as can be seen in the phase diagram (Fig. \ref{pd}).
In both rules, the average flow entirely increases with increasing $\rho_{\rm{op}}$ when $\rho_{\rm{op}}\leq\frac{1}{2}$.
As we further increase $\rho_{\rm{op}}$, \textcolor{black}{the decrease in $\bar{J}$ in the controlled phase becomes gentle, and it} can be achieved for a larger average density. 
Therefore, from a practical perspective, the most desirable $\rho_{\rm{op}}$ depends on $\bar{\rho}$ (and $\rho_{\rm{cl}}$).

In contrast with the queuing rule, the detouring rule does not stop flow even if the next node is closed by redistributing it to other open nodes.
Hence the average flow is larger than that operated under the queuing rule.
This relationship is also verified by a scaling transformation between them; $\frac{J_{\rm{q}}}{x}=J_{\rm{d}}$ with $x\leq 1$ (Eq. \ref{eq:ratio_JqJd}).

\begin{figure}[tbhp]
\includegraphics[width=90mm]{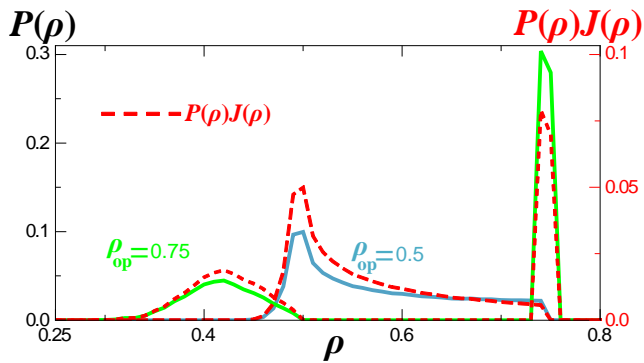}
\caption{Density distribution (solid lines) and flow distribution (red broken lines) in the controlled phase ($\bar{\rho}=0.6$).}
\label{dd}
\end{figure}

In the detouring rule, high average flow is achieved in the controlled phase. 
This can be understood by considering the density distributions of nodes, $P(\rho)$. Figure \ref{dd} shows the density distributions for  $\rho_{\rm{op}}=\rho_{\rm{cl}}=0.75$ (the base line case)
and $\rho_{\rm{op}}=0.5$. In the base line case, the distribution is bimodal, whose peaks are approximately at $\rho=0.75$ and $0.4$. In this state, once a node becomes congested ($\rho\simeq 0.75$), it cannot escape from the state
because the neighboring nodes with small density ($\rho\simeq 0.4$) send more flow to the node than the outflow from the congested node.
Thus, this density separation occurs. 
In contrast, for $\rho_{\rm{op}}=0.5$, the density distribution is unimodal and the density separation is effectively suppressed by the closure operation that discharges density in congested nodes. 
Since the flow between nodes, $J(\rho)$, takes the highest value in the middle density, collecting the density into  the middle contributes to increasing the average flow that is given by integrating $P(\rho)J(\rho)$.

\section{Discussions}
A simple control method for mitigating congestion on network transport was investigated.
We found complex behavior such as the deadlock phase transition, which is crucially important for
implementing these rules to reality. 
The control impedes density separation among nodes, thus enhancing macroscopic performance of 
the system. However, to prevent deadlock situations, one can see the trade-off relationship between
higher flow and small deadlock density. These intuitively understandable findings provide a useful overview for designing such control systems. 

The phase diagram obtained in this paper is interesting also from a physical perspective. 
The deadlock phenomenon is similar to gridlocks observed in urban traffic \cite{DaganzoUrbanGridlock} or a more simplified particle model, the BML model \cite{BML, Tadaki, Tadaki2}, 
which are caused by physical exclusion of vehicles (particles). 
In contrast, the deadlock in our model is caused by \textit{virtual exclusion} of the controlling low (that does not accept inflow). 
For this reason the phase diagram has a triple point.

As expected, the random network structure in our study enabled us to analyze and understand the system clearly. \textcolor{black}{Although this paper studies a limited range of networks,
we have confirmed that the reported characteristics are not 
significantly influenced by inhomogeneous degree distributions (in random networks and an empirical airport network in Ref. \cite{EzAir}). Different network structures would lead to changes in phase transition points (i.e., the effective traffic capacity of the system under control), which should be investigated in detail in future works. }  
We believe that they will trigger rich varieties of studies such as tolerance of transportation networks  \cite{Albert, Crucitti},  selective protection of nodes \cite{Cohen, Holme, Masuda, Takaguchi}, and the effects of spacial structure in a link \cite{Neri, EzNet,Zhang}.

\section*{acknowledgment}
TE acknowledges financial support from Japan Society for Promotion of Science Grants-in-Aid for Scientific Research (No. 13J05086).
RN and KN acknowledge financial support from Japan Society for Promotion of Science KAKENHI Grant Number 26750126 and 25287026, respectively.
We wish to thank Naoki Masuda and Daichi Yanagisawa for their valuable comments on this manuscript.


\appendix
\section{Stability of stationary flow}\label{stabcon}
Consider the state in which density is uniformly distributed in the network, i.e., $\rho_i$ $= \bar{\rho}\quad (i=1,\cdots,N)$.
By linearizing the system equation with small perturbations $(\epsilon_1,\cdots, \epsilon_N)^T$ $\equiv \bm{\epsilon}$ added to this state, 
we obtain 
\begin{equation}
\frac{d}{d t} \bm{\epsilon} = J'(\bar{\rho})\bm{M\epsilon},
\end{equation}
where matrix $\bm{M}$ is defined by $M_{ij} =A_{ji} \;(i\neq j)$ and $M_{ii}=-k$.
If all the eigenvalues of the matrix $J'(\bar{\rho})\bm{M}$ \textcolor{black}{are} not positive, the state is stable. Otherwise, a small perturbation 
grows to cause disruptions of flow. We show that the eigenvalues of $\bm{M}$ are not positive, and thus the stability of the system
is determined only by $J'(\bar{\rho}).$

Let $\bm{L}$ be a matrix defined by $\bm{L}=\bm{M} + k \bm{I}$, whose eigenvalues are $\lambda_1,\cdots,\lambda_N$. Here $\bm{I}$ is the identity matrix.
Since each element of this matrix is nonnegative, and the matrix is irreducible \cite{SC}, 
one can apply the Perron-Frobenius theorem \cite{PF}
to this matrix. Thus, we find that the spectrum radius of the matrix is $k$, i.e., $|\lambda_1|,\cdots,|\lambda_N| \leq k.$
Therefore the eigenvalues of the original matrix $\bm{M}$, $\mu_i=\lambda_i - k$, are not positive. 

In conclusion, the system state is stable if $J'(\bar{\rho})>0$ and unstable if $J'(\bar{\rho})<0.$

\section{Relationship between the queuing and detouring rules}\label{equi}
Here we apply a mean-field approximation to the links of each node as follows. We assume that $k$ directed links starting from each node are composed of $x k$ links ending in open nodes and $(1-x)k$ links ending in closed nodes. Similarly, $k$ directed links ending in each node are always composed of $x k$ links starting from open nodes and $(1-x)k$ links starting from closed nodes. 
In short, connections between links are assumed to be completely randomized, using $x$ being the ratio of open nodes.  Under this assumption, the time development of $\rho_i$ under the queuing rule obeys
\begin{eqnarray}
\frac{d \rho_i}{dt_{\rm q}} = 
\left\{ \begin{array}{ll}
- k x J(\rho_i) & \mbox{($i$: closed),} \\
\sum_{j=1}^{N} A_{ji} J(\rho_j) - k x J(\rho_i) & \mbox{($i$: open),} \\
\end{array} \right.
\label{eq:dynm_q}
\end{eqnarray}
and under the detouring rule, it obeys
\begin{eqnarray}
\frac{d \rho_i}{dt_{\rm d}} = 
\left\{ \begin{array}{ll}
- k J(\rho_i) & \mbox{($i$: closed),} \\
\sum_{j=1}^{N} A_{ji} J(\rho_j) / x - k J(\rho_i) & \mbox{($i$: open),} \\
\end{array} \right.
\label{eq:dynm_d}
\end{eqnarray}
where subscripts q and d denote the queuing and detouring rules, respectively. It is straightforward from Eqs. (\ref{eq:dynm_q}) and (\ref{eq:dynm_d}) that in the mean-field approximation, the queuing and detouring rules cause the identical time development with a scaling transformation, $x t_{\rm q}=t_{\rm d}$. 
Correspondingly, the following relationship holds for average flow rates under the two rules.
\begin{eqnarray}
\frac{\bar{J}_{\rm q}}{\bar{J}_{\rm d}} = x. 
\label{eq:ratio_JqJd}
\end{eqnarray}

\begin{figure}[tbhp]
\includegraphics[width=80mm]{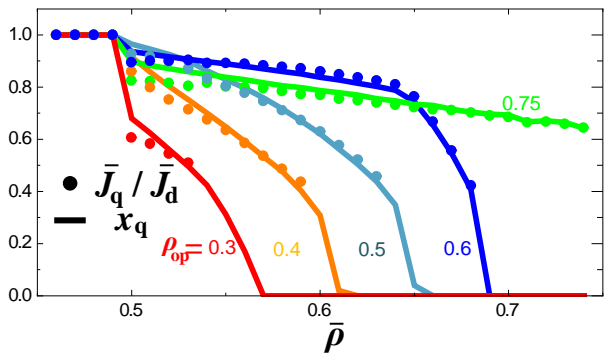}
\caption{Comparison between $\frac{J_{\rm{q}}}{J_{\rm{d}}}$ and $x_{\rm{q}}$. Initial conditions are identical to those of Fig. 4.}
\label{jjx}
\end{figure}

Figure \ref{jjx} exhibits the comparison between $\frac{\bar{J}_{\rm{q}}}{ \bar{J}_{\rm{d}}}$ and $x_{\rm{q}}$, which should be 
the same value if the two rules are equivalent. 
The agreement in the figure shows this equivalence is valid when the degree, $k$, is large.

\section{Influence of degree $k$ on system dynamics}\label{infk}
\begin{figure}[tbhp]
\includegraphics[width=45mm]{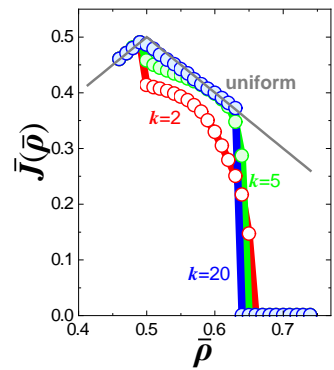}
\caption{Flow-density relationship for different degree, $k$. Here we set $\rho_{\rm{op}}=0.5$.
Other conditions are identical to those of Fig. 4(b).}
\label{ks}
\end{figure}

Figure \ref{ks} displays the flow-density relationship for different degree values, $k$.
Difference in $k$ does not influence qualitative characteristics of the system.
As $k$ increases the deadlock transition sharpens. When $k$ is small, flow is often 
completely blocked because of the small number of routes from a node. 
As a result, the average flux is suppressed, but at the same time, due to the restriction
of flow, the deadlock transition occurs at higher density.

\bibliography{basename of .bib file}

\begin{thebibliography}{00}
\bibitem{GDP}Programs that control departure time of aircrafts on the ground when a destination airport cannot accept them because of bad weather, heavy congestion or an aircraft accident.
\bibitem{SC}We are considering strongly connected networks. Normally, road networks and airport network are strongly connected, and otherwise, they can be decomposed to strongly connected subnetworks.
\end{thebibliography}

\end{document}